# Quantitative analysis of reptation of partially extended DNA in sub-30 nm nanoslits

Jia-Wei Yeh [1, 2], K. K. Sriram [1], Alessandro Taloni [3], Yeng-Long Chen[1, 4, 5], and Chia-Fu Chou[1, 6, 7,*]

[1]Institute of Physics, Academia Sinica, 128, Sec. 2, Academia Road, Nankang, Taipei 11529, Taiwan.

[2] School of Applied and Engineering Physics, Cornell University, Ithaca, NY 14850, USA.

[3] CNR-IENI, Via R. Cozzi 53, 20125 Milano, Italy.

[4] Department of Chemical Engineering, National Tsing Hua University, 101, Sec. 2, Kuang-Fu Road, Hsinchu 30013, Taiwan.

[5] Department of Physics, National Taiwan University, 1, Sec. 4, Roosevelt Road, Daan, Taipei 10617, Taiwan.

[6] Research Center for Applied Sciences, and [7] Genomics Research Center, Academia Sinica, 128, Sec. 2, Academia Road, Nankang, Taipei 11529, Taiwan.

**Abstract**

**We observed reptation of single DNA molecules in fused silica nanoslits of sub-30 nm height. The reptation behavior and the effect of confinement are quantitatively characterized using orientation correlation and transverse fluctuation analysis. We show tube-like polymer motion arises for a tense polymer under strong quasi-2D confinement and interaction with surface-passivating polyvinylpyrrolidone (PVP) molecules in nanoslits, while etching-induced device surface roughness, chip bonding materials and DNA-intercalated**

**dye-surface interaction, play minor roles. These findings have strong implications for the effect of surface modification in nanofluidic systems with potential applications for single molecule DNA analysis.**

Polymer reptation is one of the transport mechanisms for biological macromolecules in a crowded environment, such as an actin filament powered by molecular motors that snakes its way towards the center of the sarcomere [1], or a DNA molecule that wraps around histone proteins during nucleosome reposition [2]. A phenomenological approach to describe reptation dynamics is the classical tube model of Edward, de Gennes and Doi [3-6]. The main assumption is that the polymer is confined in a tube-like constraint along the polymer backbone, and its preferential motion in the tube resembles the slithering of a snake. Reptation motion has been directly observed for actin filament diffusing in an entangled actin network [7], single DNA molecule in a entangled polymer solution [8], and also for a tense DNA stretched by optical tweezers, for which the DNA transverse fluctuations are restricted into a tubelike region [9]. More recently, the reptation of stiff carbon nanotube filaments confined in a crowded environment was compared to the reptation of flexible and semi-flexible filaments [10]. Reptation motion has also been observed for a single polymer adhering on a surface, where the polymer motion is constrained to two dimensions [11-13]. To characterize reptation dynamics, the tube model must be tested by probing the trajectories of individual chains [6, 9, 14, 15]. Recent studies involving DNA motion driven by an electric field in 20 nm nanoslits has been described as biased-reptation behavior [16, 17]. Despite providing evidence of the polymer reptation dynamics, the chain trajectories and the ensuing slithering motion were not quantitatively characterized in the former experiments. In this study, we directly observe and systematically probe the trajectories of single stretched DNA polymers in nanoslits of height between 23 to 110 nm. For the first time, we provide a scaling relationship of the tube as a function of the confinement and the slit length. Most importantly, our analysis constitutes the first quantitative study of a polymer reptation in a two dimensional confined environment.

*Description of the experiment-* The experimental evidence of DNA reptation under strong slit-like confinement is provided in Figure 1a-b. Biased-reptation motion [16, 17] was induced by forcing DNA molecules into long nanoslits (25 nm in depth, 200 μm in length) using a square wave electric field. Time sequences of DNA molecules moving inside the slit clearly demonstrate its traces constrained in tube-like regions. This observation, however, only constitutes a qualitative indication of the constrained reptation dynamics. To quantitatively characterize the trajectories of single DNA molecule in nanoslits, DNA molecules were subjected to tug-of-war (TOW) and retraction scheme at the micro-nanofluidic interfaces [18], as illustrated in Figure 1d-f. For this, a T4-DNA polymer is transposed from the microchannel region into the nanoslit by applying an electric field (~2 V/cm), to overcome the entropic barrier. When one end of the polymer chain reaches the other side of the nanoslit, the field is turned off. At this point, most part of the DNA chain is in confined nanoslit region, with the coiled DNA ends protruding into the microchannel regions, and the TOW starts. The conformational entropy difference between a polymer chain confined in the nanoslit and in the microchannels induces equal and opposite entropic recoiling forces $f_{rec}$ at the two micro-nano interfaces, thereby establishing a TOW between two coiled ends of the same DNA polymer across the nanoslit (Fig. 1d). This scenario could last for minutes until one end of the DNA loses the TOW, thus entering the nanoslit at time $t_0$ from one side, retracting slowly, and eventually exiting from the other side at time $t_f$ (Fig. 1e and the movies M1, M2, and M3 in the supporting information). As demonstrated in Figure 1d, the contour of the polymer in the nanoslit remains nearly constant during the TOW. Possible causes for the reptation-like behavior include reduced transverse fluctuations due to tension acting on the DNA segment, strong surface interactions, and large surface roughness. These factors are systematically examined here using nanoslits of different height and surface coating.

*Device fabrication:* The devices were fabricated in fused silica using standard photolithography and etching methods, with the design illustrated in Figure 1c. Dry and wet etching methods were both tested for roughness influence. For dry etching of nanoslits, reactive ion etching (RIE; ANELVA DEM-451T) with a $CF_4/O_2$ mixture at 20 W was carried out; for wet etching, ten times diluted buffered hydrogen fluoride (BHF, 6:1 volume ratio of 40% $NH_4F$ in $H_2O$ and 49% HF in $H_2O$) was used. The dimensions of each nanoslit are 10 or 30 μm in width, 2, 3, 4, 5, 10, 20, 30 or 200 μm in length and 23, 28, 40, 50, 65, 110 nm (dry etching) and 25 nm (wet etching) in height were made. Nanoslits bridge inductively coupled plasma (ICP, Samco RIE-10iP) etched microchannels of 1.5×100 (depth×width) μm. Etched fused silica chip was bonded to a cover glass using two different methods. The first method uses polysilsesquioxane (PSQ)-coated cover glass subjected to oxygen plasma surface-treatment at room temperature [19]. PSQ solution is a mixture of xylene and Hardsil (Gelest) in 2:1 ratio, filtered using a syringe filter with 0.45 μm PTFE membrane (Basic Life). The second method is fusion-bonding [20, 21], which helps sealing fused silica chips to cover glass by thermal bonding at a temperature of 950 °C. In both cases, a Piranha solution (con. $H_2SO_4/H_2O_2$ in 1:1 ratio) was used to clean the substrates and cover glass in the first step. After the bonding process, sample loading reservoirs were attached to the substrate by UV-curable glue (No. 108, Norland optical adhesives). Electrical contacts were made by inserting a gold electrode in each of the reservoirs. Nanoslit surface roughness was measured by AFM (Bioscope II, Veeco) with the root-mean-squared roughness ($R_a$) around 2 nm for dry etching and 0.7 nm for wet etching. The relative roughness is typically less than 10 percent of the slit height, and does not strongly affect the DNA trajectories.

*Sample preparation:* T4 DNA (T4GT7 DNA, 166 kbp, Wako Japan) was stained with either SYBR-Gold (Invitrogen) or YOYO-1 (Invitrogen) fluorescent dye, with dye to base-pair ratio of 1:5. DNA samples were initially prepared at 0.1 μg/ml in 0.5×TBE buffer (Sigma) containing 2.5% (w/w) poly(n-vinylpyrrolidone) (PVP, 10 kDa, Sigma, used to suppress electro-osmotic flow), 30% (w/v) sucrose (J.T. Baker), and 10% (w/v) glucose (Sigma) used to increase solution viscosity thereby slowing down the dynamics of DNA molecules. The buffer viscosity was 4.1 cP, measured by a viscometer (Toki Sangyo). An oxygen scavenging system containing 0.5 %(v/v) β-mercaptoethanol (Sigma), 50 μg/ml glucose oxidase (Sigma), and 10 μg/ml catalase (Roche) was used to reduce photobleaching and help extend the observation for a few hours. The ionic strength ($I = 1.79 \times 10^{-2}$ M) of the buffer condition gives the Debye length of 1.6 nm and the effective DNA diameter ($d_{eff}$) is 10 nm [22].

*Fluorescence imaging:* Single DNA molecules were observed with a fluorescence microscopy system consisting of an inverted microscope (Leica DMI6000), 100× oil-immersion lens (Leica) with a numerical aperture of 1.4, and electron-multiplying charge coupled device camera (EM-CCD, Andor Technologies; IXon-888 or 897) with a pixel resolution of 0.13 or 0.16 μm, respectively. Images were captured at a rate of 17 frames/s. DNA trajectories were extracted from the videos by using ImageJ software (NIH) and algorithms edited by MATLAB (The Mathworks, Natick, MA). For analysis of DNA contour trajectory, information was extracted by vectorization. The fluorescence point spread function (PSF) was determined, and DNA images were iteratively deconvolved. An algorithm was written to track the fluorescence peak and intensity area by Gaussian fitting from one end of the DNA polymer to the other end for each pixel segment. In this way, the contour trajectory inside a nanoslit was determined.

*Tug-of-war scenario:* The present study allows characterization and subsequent determination of reptation dynamics in quasi-static TOW phase and during far-from-equilibrium retraction, respectively. In previous studies, reptation behavior has been identified by direct observation of DNA molecules [8, 9, 11], carbon nanotubes [10], or actin filament entangled dynamics [7, 15, 23], taking place in an imaginary "tube" that follows the molecule's or stiff filament's backbone. Our experimental setup allows quantitative determination of the "tube", by measuring the transverse fluctuation magnitude of DNA during the TOW. As a matter of fact, the "tube" can be visualized in a simple way by superimposing a sufficient number (300~2000) of contour images where the time scale of each image is around 0.1 second, as shown in Figure 2a. This observation can be quantitatively characterized with the transverse chain fluctuations defined as the root-mean-squared standard deviations of the position of DNA segment $y_i(x,t;h)$. [8, 23, 24]. $x$ is segmental position along the slit length $l_s$ $\left(x \in [0, l_s]\right)$, $i$ identifies the $i_{th}$ DNA molecule sample in slits with given $h$ and $l_s$ ($i \in [1, N_{n,l_s}]$), $N_{n,ls}$ can be between 30 or 50 molecules according to the value of $h$ and $l_s$, and $t$ is the time that the $i^{th}$ molecule spends within the slit in the TOW phase $\left(t \in \left[0, t_f^i\right]\right)$. The magnitude of fluctuations of the confined DNA around its average position is then obtained by the standard deviation

$$\sigma_i(l_s, t; h) = \sqrt{\frac{1}{l_s} \sum_{x=0}^{l_s} (y_i(x, t; h) - \langle y_i(l_s, t; h) \rangle)^2} \qquad (1)$$

where $\langle y_i(l_s, t; h) \rangle = \sum_{x=0}^{l_s} y_i(x, t; h)/l_s$. Expression (1) is then averaged with respect to the time frames collected during the TOW regime, i.e. $\sigma_i(l_s; h) = \sum_{t=1}^{t_f^i} \sigma_i(l_s, t; h)/t_f^i$, and also with respect to the ensemble of $i = N_{h,ls}$ molecules: $\sigma(l_s; h) = \sum_{i=1}^{N_{h,ls}} \sigma_i(l_s; h)/N_{h,ls}$. Therefore the former expression allows investigating the dependence of the average molecules transverse fluctuations upon the degree of confinement $h$ and the slit length $l_s$, as shown in the inset of Fig. 2b and in Figs. S3-S6 of the Supplementary material (black filled symbols). The plots display the apparent trend for which σ$(l_s; h)$ increases monotonically, with slit length $l_s$ for a given $h$. However, an accurate determination of the analytical dependence of σ$(l_s; h)$ on the confinement $h$, which is the principal aim of this analysis, is not easily inferred. This is possibly ascribable to two reasons: first, performing spatial average before temporal average could lead to different results than vice versa, although for ideal infinite chains this order is interchangeable; second, edge effects may considerably affect the overall scaling as the chain segments are more mobile close to $x=0$, $x=l_s$ (see movies M1, M2, M3) We test the first hypothesis by plotting the values of σ$(l_s; h)$ achieved by performing temporal average before spatial average (See supplementary material): Fig. S2 shows the same trends apart from a nearly constant offset. The second hypothesis was tested by defining a local spatial average, i.e. an average of the confined DNA transverse coordinate over a spatial window of length $l_x$, starting from $x_0$,

$$\langle y_i(l_x;t,h)\rangle_{x_0} = \sum_{x=x_0}^{x_0+l_x} y_i(x,t;h)\Big/l_x \qquad (2)$$

where $l_x \in [0, l_s]$, and $x_0 \in [0, l_s - l_x]$. When $l_x \equiv l_s$, one recovers $\langle y_i(l_x,t;h)\rangle_{x_0} \equiv \langle y_i(l_s,t;h)\rangle$. We then define the local transverse fluctuation by means of the standard deviation

$$\sigma_i(l_x,t;h) = \sum_{x_0=0}^{l_s-l_x}\left\{\sqrt{\sum_{x=x_0}^{x_0+l_x}\left[y_i(x,t;h) - \langle y_i(l_x,t;h)\rangle_{x_0}\right]^2 \Big/l_x}\right\}\Big/(l_s - l_x + 1) \qquad (3)$$

By averaging over time and $N_{h,ls}$ realizations, we finally obtain the mean local transverse chains fluctuation $\sigma(l_x;h)$ shown in Fig. 2b and Figs. S3-S6 of the supporting information (open symbols). The above definition enjoys a largely improved statistics for spatial windows $l_x \leq l_s$, and, on the other side, it allows a direct comparison of the local transverse fluctuations away from the edges. Fig. 2b shows an average scaling behavior of the form

$$\sigma_i(l_x,t;h) = A(h)l_x^{\xi(h)} \qquad (4)$$

The fits of the experimental curves with larger $l_x$ are reported in Fig. 2b, while the values of the coefficient $A(h)$ and of the scaling exponent $\xi(h)$ are displayed in Fig. 2c. The scaling prefactor $A(h)$ shows a monotonic increasing behavior as a function of the confinement $h$. On the other side, $\xi(h)$ from fitting is reported in the inset of Fig. 2c to be between 0.31 and 0.395 with weak dependence on $h$: setting its average value around 0.35, the scaling coefficient $A(h)$ seems to be fairly in accordance with the form $A(h) \sim h^{\xi/2}$.

The scaling expression (4) carries two important consequences. Firstly, it states that the longer the slit, the larger are the transverse fluctuations. Secondly, it corresponds to the scaling form typical to the roughness of growing surfaces [25]. In particular, the critical exponent ξ is also known as the roughness exponent, as it defines the rugosity of a surface around its mean spatial position [26]. ξ is 1/2 for Edward-Willkinson chains and directed 2D polymers [27]: the observed values 0.31-0.4 indicate that the trapped DNA fluctuates, in the $y$ direction, less than one would expect for a Rouse chain. This is certainly due to tension exerted by the recoiling forces to the free ends. As a matter of fact, smaller nanoslits augment the configurational entropy difference across the micro-nano interface, inducing higher tensile forces on the confined DNA portion. Thus, the confinement-induced tension can be reasonably considered as the only factor playing a role in suppressing transverse chain's fluctuations. This is an important finding when compared with earlier experiments and models for reptation, where the entanglement ultimately owns to the surrounding concentrated polymer solution [7, 9, 10, 15, 23]. The recoiling forces are independent of $l_s$ [18], but strongly dependent on the confinement, being $f_{rec} \sim k_B T/h$ [28]; surprisingly, the exponent ξ does not seem to vary considerably in the range of probed $h$. It appears instead that the major role played by confinement is expressed by the pre-factor $A(h)$, which indeed is absent in 1+1 dimensional models for growing interfaces.

In addition, DNA-surface interaction becomes an important issue as the degree of confinement in a nanofluidic system is increased. Expected surface interactions are (1) steric trapping from a large surface roughness [29]; (2) DNA adsorption to the PSQ-coated bonding surface; (3) DNA interaction with surface passivating short chain polymer (PVP) attached on the channel surface [16, 17], which could act like a mechanical obstacle in the slit. The surface roughness has been characterized by AFM and found to be small. We thus further investigate the other factors.

Experiments with nanoslits sealed by PSQ-bonding [19] and fusion bonding [20, 21] were used to investigate how strongly the reptation motion depends on the DNA interaction with the surface. These experiments were performed using DNA stained with two different nucleic acid staining agents YOYO-1 and SYBR-Gold, to examine any contribution from surface interaction due to the fluorescent dye. As shown in Fig. 3a, DNA transverse fluctuations during TOW were found to be similar for YOYO-1 DNA or SYBR-Gold DNA in fusion bonded or PSQ bonded nanoslits. Thus, reptation motion does not strongly depend on either the fluorescent dyes or the bonding method.

The effect of the surface-passivating PVP polymer on DNA is addressed as follows. PVP is usually added to reduce the negative surface charge of fused silica fluidic channel and the electro-osmotic flow (EOF) [17, 29-32], and earlier studies suggest that PVP may influence the movement of DNA in shallow nanoslits (~20 nm), thus causing a biased reptation [16, 17]. The 10 kDa PVP used in these experiments adsorb to the surface with an average thickness of 4 nm [33, 34]. The added concentration of 2.5 % is much higher than the saturated surface adsorption value of 0.02 %. The effective slit height is thus reduced with PVP (see schematic in the inset of Fig. 3b). The influence of PVP on the reptation motion is investigated in experiments with and without PVP, for which nanoslits of heights 28 nm and 23 nm were used respectively, thereby maintaining nearly constant effective slit height. Figure 3b shows the averaged transverse fluctuations $\sigma = 0.065 \pm 0.002$ (se) μm in nanoslits with PVP coating surface is slightly smaller than in those without PVP $\sigma = 0.07 \pm 0.003$ (se) μm. Based on these numbers, no firm conclusions can be drawn on the effects of PVP on the lateral fluctuations of the DNA molecule, it appears instead that the size of the reptation tube does remain unaffected by the presence of PVP. The smaller number of observations in experiments without PVP is due to the difficulty in driving the DNA molecules into the nanoslits, owing to the influence of surface charges of fused silica channels in absence of surface passivation [16].

*Retraction scenario:* In the TOW analysis we could characterize the size of the imaginary tube constraining the dynamics of the constrained DNA. However, this observation alone, although accurate and quantitative, does not guarantee that the polymer's dynamics corresponds to a reptating one. Thus, the reptation motion is detected and further quantified during chain retraction (when the DNA freely translocate out of the confining nanoslit), by studying the segmental tangential vector correlation of the trajectories. During the retraction process, the DNA length measured along the *x* axis can be defined as $l_i(t)$ (see Fig. 1e). Hence $x \in [0, l_i(t)]$ in the reference system for which $l_i(t_0) = l_s$ and $l_i(t_f^i) = 0$. Taking $\Delta$ = 5 pixels (0.65 or 0.8 μm respective to IXon-888 or 897 EM-CCD), we introduce the segmental tangential vector $\vec{v}_i(x,t;h)$ as the vector whose components are $[\Delta, y_i(x+\Delta,t;h) - y_i(x,t;h)]$ where $x \in [0, l_i(t)-\Delta]$. Thus, we define the tangential vector, or orientation, correlation function as

$$C_i(x,t;h) = \frac{\vec{v}_i(x,t_0;h) \cdot \vec{v}_i(x,t;h)}{|\vec{v}_i(x,t_0;h)||\vec{v}_i(x,t;h)|} \qquad (5)$$

and we report its behavior in Fig. 4. For $C_i(x,t;h) < 1$, the chain contour has changed at time *t* with respect to $t_0$. For $C_i(x,t;h) \approx 1$, the chain contour has fluctuated weakly at *t* about the $t_0$ contour. Figure 4 shows the correlation function for a single molecule (Fig. 4a) and its ensemble average (Fig. 4b) at the instances for which $l(t) = 0.9\ l_s$, $0.5\ l_s$, and $0.3\ l_s$. For $h = 28$ nm (dry etching, with PVP), the trajectory of DNA segments closely follows the backbone and the correlation function varies little with time and position (see movie M1). In contrast, there are significant fluctuations for DNA molecules in $h = 50$ and $110$ nm (dry etching, with PVP) nanoslits (see movies M2 and M3). We tested the observed behavior by varying the value of $\Delta$: our analysis reveals that, changing the length of $\vec{v}_i(x,t;h)$ in the x direction, does not affect the qualitative trend exhibited in Fig. 4.

Furthermore, Fig. 4c shows that the orientation correlation function in experiments without PVP were not as highly correlated as in experiments with PVP. This indicates that the surface-passivated PVP molecules indeed interact with the DNA molecule, and the ensuing reptation motion is strongly influenced by the presence of PVP in highly confined nanoslits. If compared with the influence that PVP has on the lateral fluctuations during TOW (Fig. 3b), we can conclude that the effects of PVP during the non-equilibrium retracting dynamics are much stronger. This can be explained by the fact that PVP acts as a surrounding entanglement medium (polymer concentration) through the entire reptation process, which is indeed a dynamical event. PVP's presence although does not clearly affect the size of the tube, which instead is a stationary quantity that depends only on confinement-induced entropic tensile forces.

In analogy with the analysis performed in the TOW phase, we study the effects of PSQ-bonding and fusion bonding on the retraction phase, achieving no experimental evidence that they hinder or promote the DNA reptation (see Fig. S7).

Similar results were inferred from experiments with different DNA staining agents, YOYO-1 and SYBR-Gold (Fig. S7)

Finally, it may be of interest to compare the prior results of free diffusion of DNA polymer in sub-30 nm nanoslits [35-37]. The strong confinement is known to affect the DNA segmental correlation length [38, 39], and the diffusivity measurement attributed the relation to the "rod-like" chain conformation as Odijk's argument [35, 40]. More relevant, the significant difference between the diffusivities measured along the x- and y- axis in 24 nm deep nanoslit has been dug out [37]. The origin of which could be understood due to the interaction of surface passivating PVP polymer we demonstrated here.

*Conclusions:* In this Letter we have, for the first time, provided the quantitative characterization of DNA reptation in strong confinement conditions. This quantitative analysis, obtained by detecting the DNA transverse fluctuations in nanoslits during the equilibrium phase (TOW) and the orientation correlation of the subsequent out-of-equilibrium retraction process, shows that the strong confinement considerably suppresses the DNA freedom in the transverse direction, thus leading to reptation-like motion in the dynamical phase. Indeed, from sub-Kuhn length nanoslit confinement [18, 28] (50~100 nm) to the unconfined free solution [24, 41], the elongated DNA polymer has increasing degree of transverse fluctuations and segmental orientations. On the other hand, in strongly confining nanoslits ($\leq 30$ nm), DNA molecules are restricted in a tube-like region. Our study provides, for the first time, a quantitative characterization of the tube with confinement, by means of scaling relations whose critical exponents and prefactors depend on the slit height. Moreover, we have confirmed that the presence of PVP in solution significantly alters the reptation dynamics, playing the active role of an entanglement medium. On the other side, the motion appears not to be influenced by the bonding surfaces or the fluorescent staining dyes. These findings help clarify the enforced reptation behavior observed in nanoslits from recent reports [16, 17, 29]. In a 3D gel environment, DNA molecules migrate through the entanglement polymer concentration by reptation and form a tube-like behavior. In this experiment, we have provided the evidence that the DNA reptation in 2D slit-like nanochannel is not only due to the confinement induced tensile forces, but also a major role of PVP by acting like a 2D gel matrix. Our work may lead to an improved device design, using simple 2D nanoslits to achieve high degree of DNA extension, and low degree of thermal fluctuation for high-resolution single molecule analysis [42-45].

**Acknowledgements**

We thank Profs. Andreas Erbe, Cheng-Hung Chang and David Newquist for their critical review of the manuscript and insightful suggestions, Drs. Po-Keng Lin and Sheng-Qin Wang for helpful discussions, Wei-Chiao Lai for AFM measurement, and technical support from the Academia Sinica (AS) Nano Core Facilities. This work was supported by the AS Program on Nano Science and Technology and AS Thematic Projects [AS-97-FP-M02, AS-103-TP-A01] (to C.F.C.) and by the Ministry of Science and Technology (ROC) [102–2112-M-001-005-MY3, 103–2923-M-001-007-MY3] (to C.F.C.).

**Figures and Captions:**

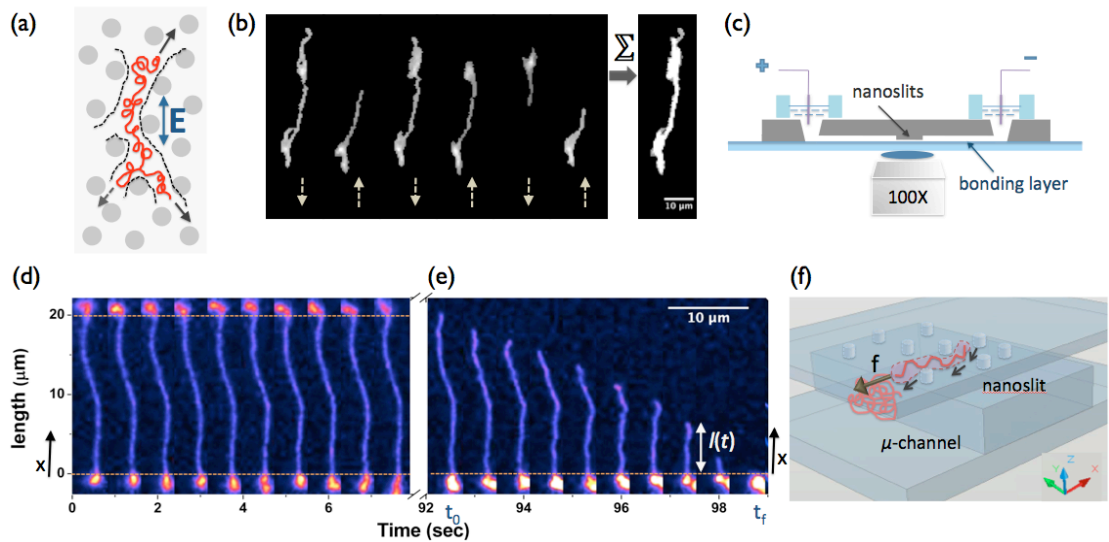

**Figure 1.** (a) Schematic showing reptation of entangled polymer driven by an electrical field in nanoslits, with gray circles representing fictitious pillars to illustrate polymer entanglement. (b) Time sequence images and superposition of single molecule DNA trajectories in 25 nm wet-etched nanoslits. The images are separated by 0.135 sec. Applied field: 1333 V/m, 1 Hz AC; White arrow indicates the field direction. (c) Schematic of our micro/nano-fluidic device setup. (d) A DNA molecule during TOW [18]. Dotted lines represent micro-nano-interfaces. Upper and lower parts are microchannels; the middle part is the nanoslit region. (e) A DNA molecule retracting out of the nanoslit. The time interval between frames is 0.7 sec. The slit length and height are 20 μm and 28 nm, respectively. Translocation process takes place between $t_0$ and $t_f$, the time of complete translocation through the slit, being the retracting length $l(t)$. (f) Schematic of DNA reptation in nanoslit. Fictitious pillars are drawn to illustrate polymer entanglement.

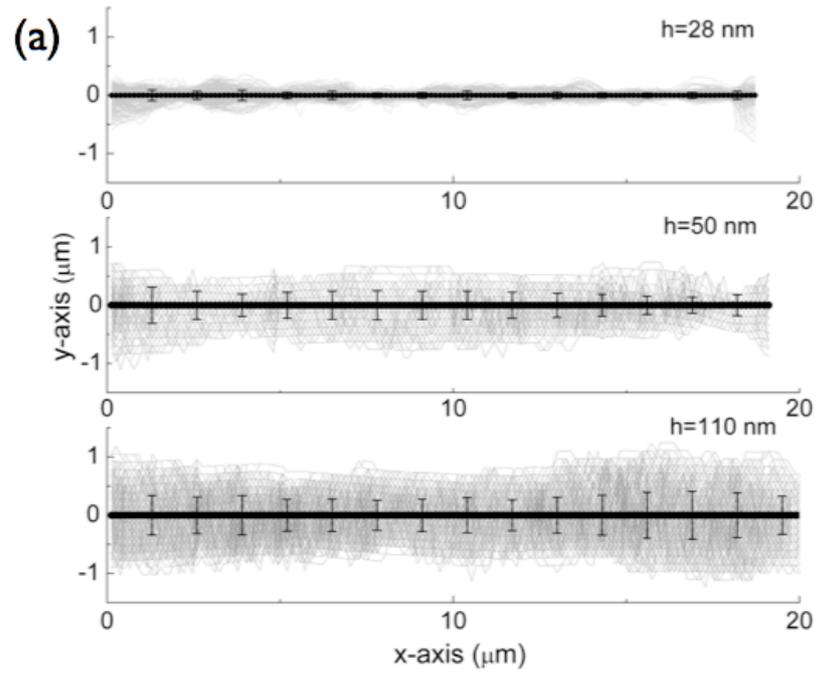

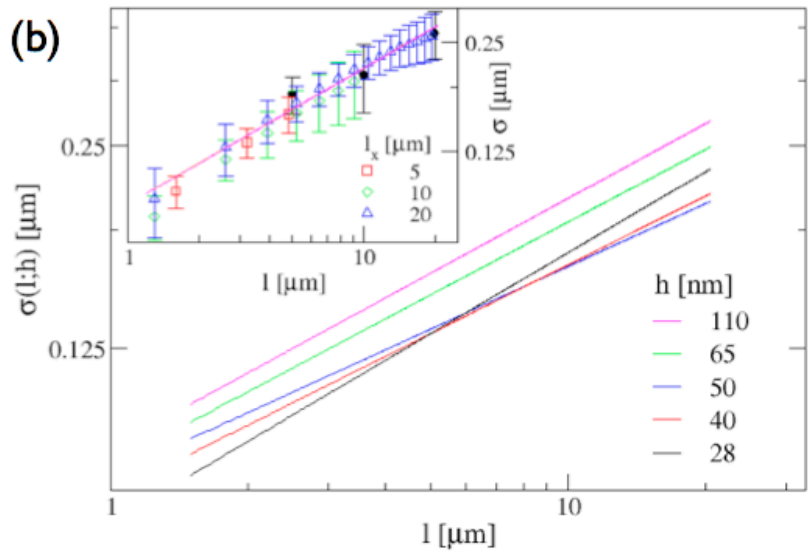

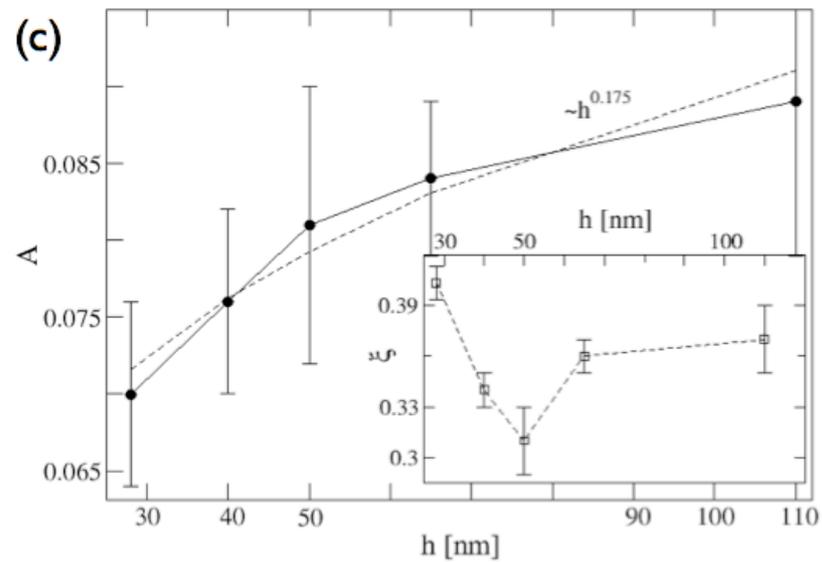

**Figure 2.** (a) Superimposed frames (separated by 0.1 s) of the DNA contour to determine the transverse fluctuation of DNA trajectories during TOW. The gray lines are 500 to 4000 trajectory traces of a DNA for $h$ = 28, 50, and 110 nm (dry etching, with PVP) and the solid lines represent the corresponding averaged curves. (b) Average fluctuation vs. the longitudinal lengths $l$ for $h$ = 28, 40, 50, 65 and 110 nm (dry etching, with PVP). The solid lines represent the best fits of the transverse local fluctuation defined in Eq. (4) ($l = l_x$, see the inset and Figs. S3-S6). Inset: the filled black symbols show the result of the global fluctuations as a function of $l_s$ for $h$ = 110 nm ($l = l_s$). The local roughness defined by equation (3) allows analyzing the scaling of the transverse fluctuations for values of $l_x \leq l_s$, shown as open symbols ($l = l_x$). The best fit of the numerical curves according to expression (4) is displayed in magenta and corresponds to the top curve in the main panel. (c) Scaling prefactor $A(h)$ and roughness exponent $\xi(h)$ (inset) as a function of $h$. The value of $h$ seems to affect prefactor $A$ considerably, showing an apparent reduction of the tube corresponding to the reptation dynamics, as the confinement increases. The scaling exponent $\xi$ shows decay in correspondence of the DNA persistent length (~ 50 nm).

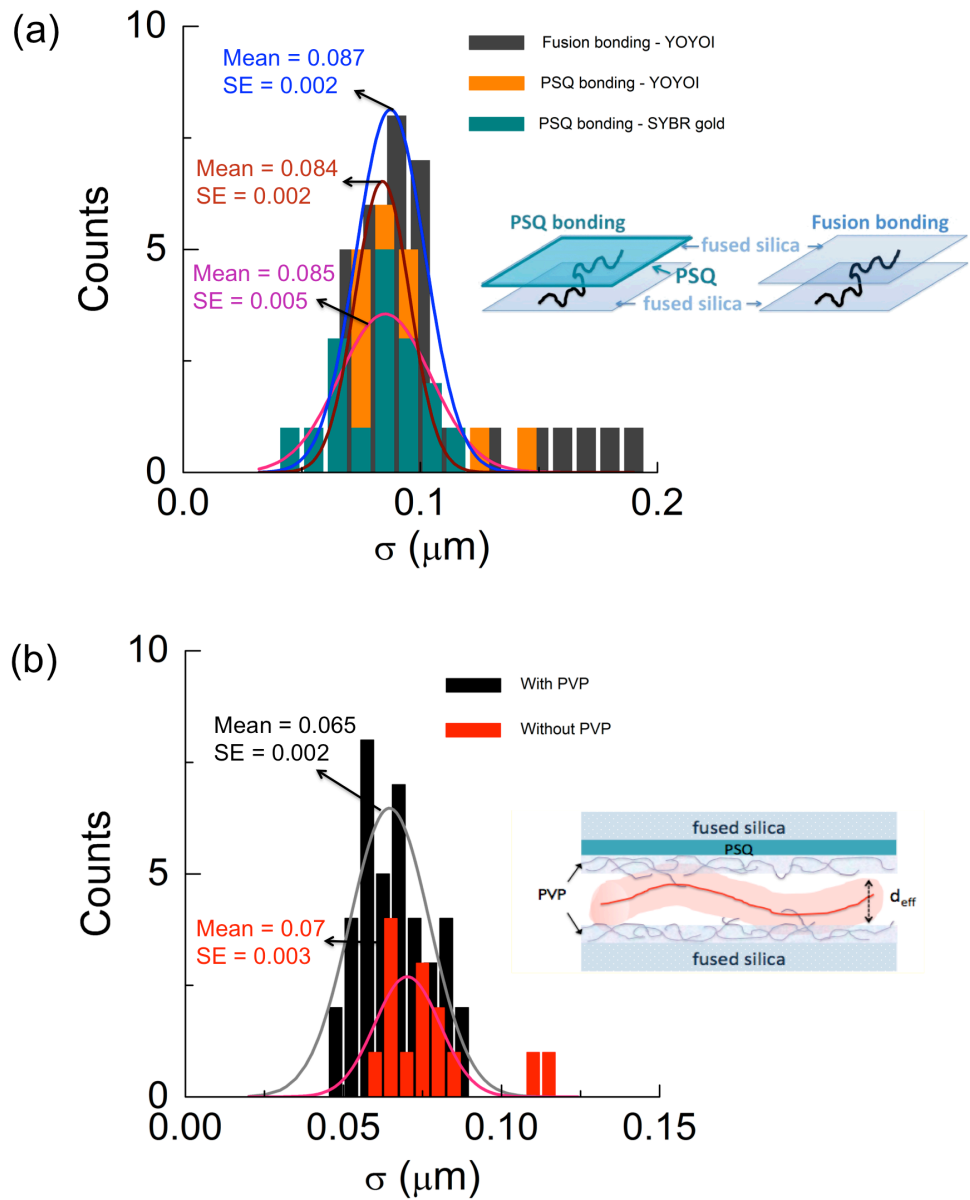

**Figure 3.** (a) Histogram of transverse fluctuation for YOYO-1 and SYBR-Gold dye labeled DNA in PSQ- and fusion-bonded nanoslits of 28 nm height (dry etching, with PVP) and 20 μm length. (Numbers of molecules: PSQ/YOYO-1: 19; FUSION/YOYO-1: 20; PSQ/SYBR-Gold: 31) (b) Histogram of transverse fluctuation for DNA in PSQ-bonded nanoslits, 10 μm length, of heights 28 nm (dry etching, with PVP) and 23 nm (dry etching, w/o PVP), respectively, but the effective slit height is the same in both cases. Inset shows illustration of DNA (red line) in the nanoslit. The effective diameter ($d_{eff}$) is ≈ 10 nm; thickness of PVP layer (purple line with transparent blue) is 4 nm on each side. (Numbers of molecules: w/ PVP: 45; w/o PVP: 15)

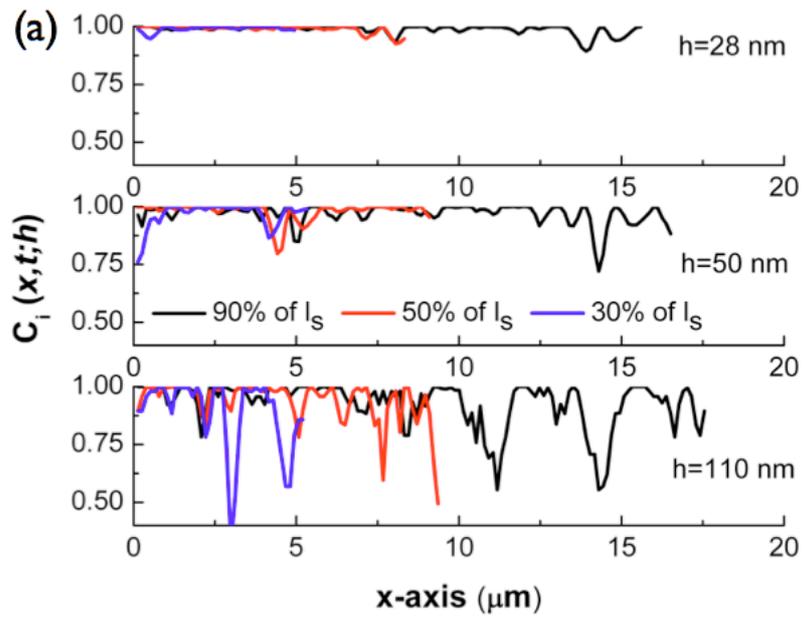
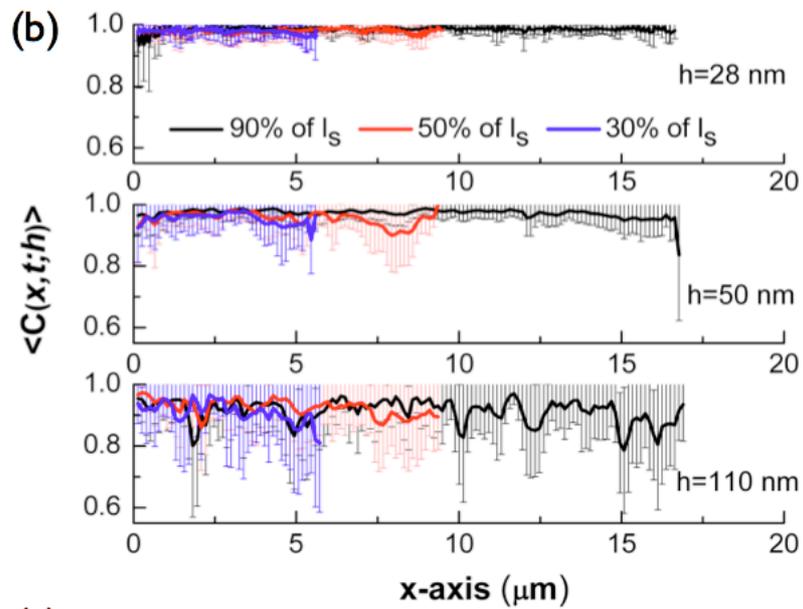
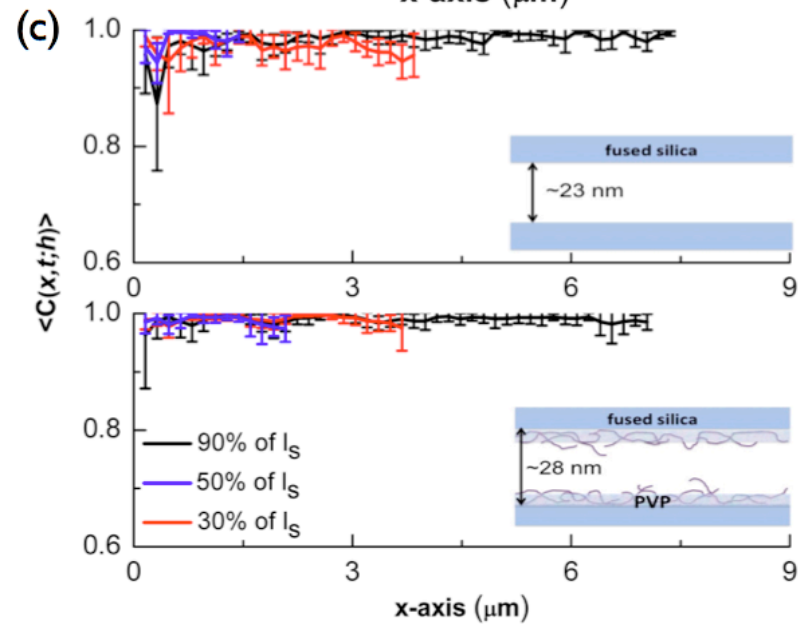

**Figure 4.** (a) The tangential vector, or orientation, correlation function $C_i(x,t;h)$, for $h$ = 28, 50, and 110 nm (dry etching, with PVP) for slit length 20 μm for segments along the DNA backbone at times where the projected chain along the $x$ axis is equal to 90% (black line), 50% (red line), and 30% (blue line) of the nanoslit length ($l_s$) during retraction recoiling. (b) The ensemble average of orientation correlation $\langle C(x,t;h) \rangle$ for $h$ = 28, 50, and 110 nm (dry etching, with PVP) for slit length 20 μm for segments along the DNA backbone at times where the projected chain along the $x$ axis is equal to 90% (black line), 50% (red line), and 30% (blue line) of the nanoslit length ($l_s$) during retraction recoiling. (c) Ensemble average of orientation correlation, $\langle C(x,t;h) \rangle$, for experiments done with and without PVP (~10 molecules for each case) in PSQ bonded nanoslits at the 90% (blue line), 50% (red line), and 30% (black line) of nanoslit length during retraction recoiling.